\documentclass[epj]{svjour}
\usepackage{graphics}
\usepackage[dvips]{graphicx,color}
\usepackage[body={16cm,22cm},bottom=1.5cm]{geometry}
\usepackage{amsmath}
\usepackage{psfrag}
\usepackage{latexsym}
\usepackage{amsfonts}
\begin{document}
\title{On the study of local stress rearrangements
during quasistatic plastic shear of a model glass: do local stress components contain enough information?}
\author{Michel Tsamados \inst{1},
Anne Tanguy \inst{1},
Fabien L\'eonforte\inst{2},
Jean-Louis Barrat\inst{1}
}                     
%
%
\institute{Universit\'e de Lyon; Univ. Lyon I, Laboratoire de
Physique de la Mati\`ere Condens\'ee et Nanostructures; CNRS, UMR
5586, 69622 Villeurbanne, France \and INSA Lyon, MATEIS, 69621 Villeurbanne, France}
\date{Received: date / Revised version: date}
%
\abstract{
We present a numerical study of the mechanical response of a 2D
Lennard-Jones amorphous solid under steady quasistatic and athermal
shear. We focus here on the evolution of local stress components.
While the local stress is usually taken as an order parameter in the
description of the rheological behaviour of complex fluids, and for
plasticity in glasses, we show here that the knowledge of local
stresses is not sufficient for a complete description of the
plastic behaviour of our system. The distribution of local stresses
can be approximately described as resulting from the sum of
localized quadrupolar events with an exponential distribution of
amplitudes. However, we show that the position of the center of the
quadrupoles is not related to any special evolution of the local
stress, but must be described by another variable.
\PACS{
      {PACS-key}{discribing text of that key}   \and
      {PACS-key}{discribing text of that key}
     } 
} 
\authorrunning{M. Tsamados \it{et al}}
\titlerunning{Stress rearrangements during quasistatic plastic shear }
\maketitle
\section{Introduction}
While the mechanical plastic response of crystalline solids is now
well described in terms of elementary defects called dislocations,
the plastic response of amorphous solids is not so well understood,
 and has been the object of numerous studies in the past two decades.
Most theoretical works~\cite{Falk1998,Falk1999,Argon1995,Bulatov1994,Bulatov1994a,Bulatov1994b,Sollich1997,Sollich1998,Roux2002,Baret2002,Picard2005,Rottler2003,Wyart2005,Wyart2005b,Demko2005,Maloney2004a,Maloney2004,Maloney2006,Lemaitre2007}, on the forced dynamics of glassy materials assume that the irreversible evolution of glasses occurs
through localized, irreversible events.
Such events have been well identified in simulations of 2D models
~\cite{Falk1998} and are referred to as shear transformation zones
(STZ) or quadrupolar events~\cite{Maloney2006,Lemaitre2007,Lemaitre2007a,Tanguy2006}
and are somewhat similar to T1 events in the rheology of 2D foams~\cite{Dennin2004}.

 These localized and dissipative events should correspond to the ``soft
mode" of the material ~\cite{Maloney2006,Lemaitre2007,Malendro1998,Lacks2004}),
which is the first mode to become unstable, when the system reaches a limit of mechanical
instability.
However, the local and intermittent nature of the plastic events
complicates the identification of the corresponding order and
control parameters. Experimental studies on disordered materials,
far below the glass transition temperature
\cite{Dennin2004,Besseling2007,Lauridsen2002,Howell1999,Tewari1999,Majmudar2005,Kolb2004,Debregeas2001,Kabla2003},
support the existence of a collective behaviour of localized
rearrangements leading to a strongly heterogeneous mechanical
response as shown in the existence of shear bands in the macroscopic
plasticity of such systems. It has also been shown experimentally
~\cite{Kabla2003,Cantat2006,Marmottant2007} as well as numerically
~\cite{Falk1998,Maloney2006,Tanguy2006,Durian1995,Varnik2003,Bouchbinder2007} while
studying the stress-strain response of amorphous materials that, in
the so called plastic regime, these events lead to an intermittent
release of the macroscopic stress in the material. The understanding
of the role played by local stresses in the macroscopic mechanical
behaviour of such systems is thus very important. Since these
systems are known to have a wide distribution of local stresses,
mesoscopic models
\cite{Sollich1997,Baret2002,Picard2005,Hebraud1998} propose
precisely to relate the existence of shear bands to the local
rearrangements due to local thresholds and activated dynamics in the
local stress itself. In these models, the collective effects would
thus be due to long-range elastic couplings in solid systems, and
the irreversible behaviour would be associated  with local stress
thresholds, and avalanche dynamics.

In the present paper, we provide an extensive study of the local
stress dynamics in the quasi-static plastic response of a model
glass at very low temperature, and in the plastic flow regime. The
system we have studied is a 2D Lennard-Jones glass, already studied
previously for understanding the heterogeneous mechanical response
in the elastic regime, far below plastic flow
~\cite{Wittmer2002,Tanguy2002,Leonforte2004,Leonforte2005,Tanguy2006}.
In the first part of this paper, we discuss the  evolution of the
local stresses (local pressure, deviatoric stress) in the plastic
flow regime, within the framework of dynamical heterogeneities
~\cite{Toninelli2005}. We show the existence of a finite size of
cooperativity appearing in the plastic flow regime, even in the
limit of zero temperature and quasi-static shear. In the second part
of this paper, we provide a detailed analysis of the full
statistical properties of the stress components (pressure, shear
stress). We show that the temporal correlations shown in the local
particle rearrangements for small imposed strain
~\cite{Tanguy2006,Lemaitre2007} are absent in the average evolution
of local stress components. We then show in the third part of this
paper, that the statistical evolution of stress components can be
described as resulting simply from a sum of localized quadrupolar
rearrangements. This description allows us to propose a simple
equation of evolution for the local stress depending unfortunately
on a local yet unknown criterion for instability. In the last part,
we show that this local criterion for instability is not related to
any stress threshold but could be related to a different variable,
such as the local stress increments.

\section{Dynamical heterogeneity}
\subsection {Two points correlation function}

The system we study numerically has already been described
extensively in previous articles~\cite{Tanguy2002,Leonforte2004,Leonforte2005,Tanguy2006}).
It is a 2D glass obtained by quenching very quickly a liquid sample
obtained by molecular dynamics simulations, and made of polydisperse
particles interacting through a Lennard-Jones potential. In the
following, all quantities will be expressed in Lennard-Jones units
($u_{LJ}$), that is an energy $\epsilon_{LJ} = 1 (\approx 0.1 eV)$, a
distance $\sigma_{LJ} = 1 (\approx 2 \AA)$ and mass $m = 1 (\approx 10
g.mol^{-1})$. The number of particles per unit surface is $\rho =
0.925$, corresponding to very small pressure $P \approx 0.2$ at
rest. The system is made of at least $N=10 000$ particles,
corresponding to a square with a linear size of $104$ interatomic
distances. Larger systems have also been considered, especially in
this paper a system of $N=23 125$ particles, corresponding t a
rectangular box of size $50\times 500$. These large system sizes are
necessary in order to avoid strong finite size effects. The
quenching starts from the liquid state, at a temperature $T=2$. Then
the system is relaxed using MD to a given temperature of
$1-0.5-0.1-0.05-0.01-0.005-0.001$, where it is aged during $1 000$
unit times. The final step is a relaxation at zero temperature,
using the Conjugate Gradient Method in order to get an instantaneous
minimization of the potential energy. The system being at zero
temperature, it is sheared quasi-statically, by imposing successive
steps of global shear strain $\delta\epsilon = 5.10^{-5}$ on the
walls. After each step, the system is relaxed into its nearest
equilibrium position, thus ensuring that the applied mechanical
deformation is quasi-static. The local stress components are
computed on each particle $i$ by using the usual Irving-Kirkwood
formula (\cite{Irving1950})
$$
\nonumber
\sigma_{\alpha\beta, i}\equiv -1/V_i.\Sigma_j t_{ij} d_\alpha.d_\beta.r_{ij}
$$
$V_i$ being the volume of the Voronoi cell, $t_{ij}$ the amplitude
of the interacting force between particles $i$ and $j$, $r_{ij}$ the
distance between particles and $\overrightarrow d$ the unit vector
of the bond. The macroscopic stresses are obtained   by averaging
the local stress components  over the whole system, or alternatively
by measuring the force per unit length acting on the walls along
appropriate directions. In Fig.1, we show the (averaged) macroscopic
shear stress as a function of the total strain applied to the
system. The global shear stress has an intermittent behaviour, with
an alternance of small and large jumps, giving rise to negative
slopes that are the signature of dissipative events. Note that the
distribution (figure 2) of these incremental stresses is peaked
around the value given by the elastic response with an elastic shear
modulus $\mu \approx 11.7$ (\cite{Tanguy2002}) and shows around this
value softer, as well as more rigid, steps.

In order to characterize the evolution of the stress in the flowing
regime, we first compute the two points correlation function of the
local stress components acting on each particle inside the sample.
The correlation function is ``Lagrangian" in the sense that the
stress is followed on a given particle as the system is sheared. In
the plastic flow regime, we expect the system reaches a stationary
steady state with time-translational invariant (TTI) observables and
correlations. Note that in quasistatic simulation, time does not
appear explicitly and must be interpreted in  the following as the
number of steps $n$ of elementary shear strain $\delta\epsilon$
applied to the system. The total imposed strain is thus
$\epsilon=n.\delta\epsilon$. In order to study the stress
fluctuations correlation function, we introduce the notation
$\rho(\overrightarrow{r},n)\equiv\sigma(\overrightarrow{r},n)-\overline{\sigma(n)}$
where $\overline{A}$ denotes a spatial average and
$\overrightarrow{r}$ is the position of the particle; $\sigma$ is a
component of the stress tensor, namely $\sigma\equiv\sigma_{dev}$
where $\sigma_{dev}\equiv\sigma_1-\sigma_2$, $\sigma_1$ and
$\sigma_2$ being the eigenvalues of the local 2x2 stress tensor; or
$\sigma\equiv\sigma_{p}$ where $\sigma_p=-(\sigma_1+\sigma_2) / 2$
is the local pressure; or $\sigma\equiv\sigma_{xy}$ is the local
shear stress. The two-time autocorrelation function of the stress
fluctuations writes

\begin{equation}
C(\Delta n,n)=\overline{\rho(\overrightarrow{r},n+\Delta n)\rho(\overrightarrow{r},n)},
\end{equation}
and
\begin{eqnarray}
\label{eqn:moytemporelle}
C(\Delta n)&=&\langle C(\Delta n,n)\rangle\nonumber\\
&=&\overline{\langle\rho(\overrightarrow{r},n+\Delta n)\rho(\overrightarrow{r},n)\rangle}\nonumber\\
&\approx& C(\Delta n,n),
\end{eqnarray}
where the last equality assumes TTI.

Numerically we have computed the stress components (as mentioned
before from the Irving-Kirkwood definition, denoted by
$\sigma_{i}(n)$) on each particle of the system. The discrete
version of the above equation writes

\begin{equation}
\label{eqn:moytemporelle}
C(\Delta n)=\frac{1}{N}\langle\sum_{i} \rho_{i}(n+\Delta n)\rho_{i}(n)\rangle.
\end{equation} where $\rho_{i}(n)=\sigma_{i}(n)-\overline{\sigma(n)}$.
We have computed this quantity in the plastic flow regime (i.e. for
external shear deformation $\epsilon$ greater than $1.5\%$), where
the average total stress saturates and fluctuates around a given
average value $\sigma_Y$. We checked that the correlations are
invariant under time translation and in Fig. 3 we plot the
autocorrelation function for the shear component of the stress, on a
lin-log scale and normalized to 1 for $\Delta n=0$. The behaviour
can be decomposed into two successive relaxations. It starts with a
first exponential decay at small deformation with a characteristic
strain of $3.1\%$, this decrease to a plateau is analogous to the
$\beta-relaxation$ in supercooled liquids, and is followed by an
apparent stretched exponential decay. A similar
Kohlrausch-Williams-Watts ~\cite{Kohlrausch1863,Williams1970}
is usually observed in supercooled liquids and ageing glassy
systems, with a correlation that evolves as $C(t)=A
\exp(-(t/\tau)^{\beta})$ with $\beta\leq 1$. This behaviour is
approximately recovered in our systems at  large deformations (see
inset of Fig. 3), with the exponent $\beta=0.7$. For small strains,
the relaxation is strictly exponential. Note that the large strains
stretched exponential decay could also be seen as the best fit for
an intermediate regime between two successive exponential decays
(see the fit by an exponential decay with characteristic strain
$\epsilon\approx 10\%$ at very large strains in Fig. 3). In any
case, the presence of dissipative events leads to the relaxation of
local stress at large deformations, above a few percents of strain.

\subsection{Four points correlation function or cooperativity number}
In the recent years, the quest for a dynamical cooperativity length
associated with the slowing down of  the dynamics of supercooled
liquids  ~\cite{Toninelli2005,Doliwa2000,Lacevic2003a} has led to
the development of new statistical tools to characterize such
``dynamical heterogeneities".  In our previous paper
\cite{Tanguy2006} we showed that the motion of the particules in our
system under shear is highly non trivial. It shows a background of
heterogeneous motion even in the elastic regime at very low
temperature; for larger applied strains, in the plastic regime, it
shows zones of very high mobility located in the vicinity of
elementary shear bands and of localized quadrupolar events, and
similarly zones of low mobility far from these irreversible events
(the local displacement field can vary by many orders of
magnitudes). This type of behaviour in the mechanical response of
glass, even at zero temperature, can be seen as some kind of
dynamical heterogeneity (while the ``dynamics" here is overdamped)
and it is of great interest to quantify a cooperativity degree or a
cooperative length scale. To do so, Berthier et al.
~\cite{Toninelli2005} have proposed to look at the so called
$\chi_{4}$ four point correlation function. An other possible
observable is the cooperativity number introduced by Doliwa and
Heuer ~\cite{Doliwa2000}, that estimates the spatial fluctuations of
the two-point correlation function introduced previously. Following
reference  ~\cite{Doliwa2000} the definition we use here is
\begin{equation}
\label{eqn:Ncoop} N_{coop}\equiv\frac{Var[\Sigma X_{i}]}{\Sigma
Var[X_{i}]}=\dfrac{ \langle \lbrace \sum_{i} X_{i}-\langle\sum_{i}
X_{i} \rangle \rbrace ^{2}\rangle }{ \sum_{i} \lbrace\langle
X_{i}^{2}\rangle - \langle X_{i}\rangle^{2}\rbrace},
\end{equation}
where
$X_{i}(t',t)=(\sigma_{i}(t+t')-\overline{\sigma_{i}(t+t')})(\sigma_{i}(t)-
\overline{\sigma_{i}(t)})$ is a dynamical quantity associated with
every particle $i$, and computed on various components $\sigma$ of
the stress tensor. In Eq.(4), one shows easily that in the case of
totally uncorrelated variables $X_{i}$  ($\langle
X_{i}X_{j}\rangle=0$), one gets $N_{coop}=1$, but in the opposite if
the correlation is complete ($X_{i}=X_{j}$) one gets $N_{coop}=N$.
Finally in the case of $L$ independent groups each comprising $M$
identical variables and of zero mean ($N=LM$) one gets

\begin{eqnarray}
\label{eqn:NcoopM}
N_{coop}&=&1+\frac{\sum_{i\neq j}\langle X_{i}X_{j}\rangle}{\sum\langle X_{i}^{2}\rangle}\nonumber\\
&=&1+\frac{1}{\sum\langle X_{i}^{2}\rangle}\sum_{i=1}^{N}\sum_{j=1}^{M-1}\langle X_{i}^{2}\rangle=M.
\end{eqnarray}

The cooperativity number is obtained by averaging over the origins
$t$ and remains a function of time $t'$ ($N_{coop}(t')$). Note that
this function does not contain any useful information on the
heterogeneous motion in the linear elastic regime, since in this
case, the amplitude of the stress on each particle increases
linearly with time. The cooperativity number is thus constant in
this regime. A maximum can be reached only in the non-linear regime,
like in the plastic flow regime here, or when thermal activation
starts playing a role. In Fig. 4 we have plotted this function
computed on all particles located at a distance of a few (typically
10) particle diameters from the boundaries, and for different
components of the stress tensor. All the stress components have
approximately the same contribution to the cooperativity number,
with a maximum of a few tens at an imposed strain of a few percents
($\approx 4 \%$). As described before, the maximum is a measure of
the maximum number of particles on which the local stress evolves in
a correlated manner. In our case, it corresponds to an ensemble of
$6 \times 6$ to $7 \times 7$ particles where the stress components
evolve in a cooperative way. The strain at which this maximum is
reached is about $4$ percents, that is beyond the plastic threshold,
and close to the characteristic strain obtained from the two points
correlation function $C$. This picture corresponds to the
description of localized plastic events, that we have made already
in a preceding paper~\cite{Tanguy2006}. In the elastic regime (as
in elastic branches present in the plastic flow plateau~\cite{Tanguy2006}),
the cooperative evolution of stress can be more important, but
does not contribute significantly to the cooperativity number.

Fig. 4 shows that there are strong boundary effects measurable
on this quantity. Indeed,  the cooperativity number is much higher
in the vicinity of the walls ($\approx 100$) corresponding to a
layer of one particle diameter along the entire system size
($L_x=104$). Finally, note that this maximum cooperativity number is
finite in the quasi-static limit studied here, suggesting a
saturation at very low shear rate.

\section{Statistical analysis of the local shear stress}
In the previous section we have studied the average evolution of the stress
components by looking at averaged quantities like the
number of cooperativity, or the two-point correlation function. We
will now examine in details the full histogram of stress changes.
The results are shown here for the shear stress component,
but we have checked that all the following results are also valid
for other components (pressure and deviatoric stress). In figure 5
we show the typical evolution of the shear stress component on a
given particle up to a total strain of $25\%$\footnote{In a previous
paper we have analyzed in a similar way the motion of a particle in
the system under shear (see Ref. \cite{Tanguy2006})}. The variations
of stresses are larger than expected if they would be Brownian. In
order to analyze this evolution, we have plotted in figure 6 the
distribution $P(\Delta \sigma_{xy},\Delta n)$ of the shear stress
increments
\begin{eqnarray}
\Delta \sigma_{xy}(\Delta n,n)\equiv &(&\sigma_{xy}
(n+\Delta n)-\overline\sigma_{xy}(n+\Delta
n))\nonumber\\&-&(\sigma_{xy}(n)-\overline\sigma_{xy}(n))
\end{eqnarray}
for various numbers $\Delta n$ of incremental shear steps ($\Delta
n=1,2,4,8,16...$), averaged over the origins $n$ and over the entire
system. The distributions $P(\Delta \sigma_{xy},\Delta n)$ are all
symmetric, with zero mean. For $\Delta n\rightarrow \infty$ one
recovers a gaussian distribution, which is consistent with the
central limit theorem. However, at small imposed strains, these
distributions are not gaussian as  would be the case for a Brownian
evolution. A finer analysis of the distribution of the elementary
increments $P(\Delta \sigma_{xy},\Delta n=1)$ (see fig 6b and c)
shows three zones. At small incremental stress jumps
$\Delta\sigma_{xy}$, we can see a plateau of approximately constant
probability, whose width evolves inversely proportional to the
volume $V$ of the sample, followed by an apparent power-law decay,
in a zone of approximately three decades for $100/V
\leq\Delta\sigma_{xy}\leq10$ where $P(\Delta \sigma_{xy},\Delta n=1)
\propto 1/{\Delta\sigma_{xy}^{\alpha+1}}$, and concluded by an
exponential cut-off independent on the system size (characteristic
shear stress $\Delta\sigma_{xy}\approx 1.4$). This upper cut-off
allows for a finite variance of the local stress evolution. In the
absence of any temporal correlations, the entire process can thus be
described by the central limit theorem, with a scale invariant
distribution on a finite interval. In this intermediate stress
range, the probability density function (hereafter, pdf) $P(\Delta
\sigma_{xy},\Delta n)$ is well reproduced by the scale invariant
relation

\begin{eqnarray}
P(\Delta \sigma_{xy},\Delta n)&=&\Delta n^{-H}f(\frac{\vert\Delta \sigma_{xy}\vert}{\Delta n^{H}})
\nonumber\\or\nonumber\\
P(\Delta \sigma_{xy},\Delta\epsilon)&=&\Delta\epsilon^{-H}f(\frac{\vert\Delta \sigma_{xy}\vert}{\Delta \epsilon^{H}})
\end{eqnarray}
with
\begin{equation}
f(u)\propto \left\{
     \begin{array}{cc}
        u^{0}                    & \mbox{, for}      \  u \ll 1 \\
        u^{-\alpha-1}            & \mbox{, for}      \  1\ll u\ll cste.\Delta n^{-H} \\
     \end{array}
   \right\}
\end{equation}

In the intermediate stress range, the process can thus be considered
as self-similar. Fig 7(a) illustrates this scaling with a good
superposition of the distributions for $\alpha = 0.7$ and
$H\simeq\frac{1}{\alpha}$ (only for not to large $\Delta n$ since
for very large $\Delta n$ the upper cut-off discussed before
contributes significantly to the resulting distribution). The
exponent $\alpha$ describes the algebraic (slow) asymptotical decay
of the distribution of the incremental jumps
$P(\Delta\sigma=s,\Delta n=1)\propto s^{-\alpha-1}$ (as shown in fig
6(a)). The exponent H is related to the evolution of the stress
jumps as a function of the applied shear strain $\Delta n$. It
characterizes the $\Delta n$-dependence of the crossover between a
regime of approximately uniform probability
(for $\vert\Delta\sigma\vert\ll \Delta n^H$) and the
power law regime (for $\vert\Delta\sigma\vert\gg \Delta n^H$).
The coefficient H also accounts for possible temporal statistical
correlation between jumps. As stated
in Taqqu et al. ~\cite{Taqqu} the only non degenerate
$\alpha-stable$ self similar processes with stationary increments
that verify $H=\frac{1}{\alpha}$ and where $0<\alpha<1$ are the
$\alpha-stable$ L\'evy motions. As described above, the evolution of
the shear stress is thus of L\'evy flight type, but only in the
intermediate stress (and applied strain) range. The L\'evy flight
evolution implies by definition, first that there is no temporal
correlation between local stress jumps during the shear of the
sample, second that the variance of elementary changes is infinite
(as long as the exponentiel cut-off is neglected)\footnote{In the
case of Brownian motion the properties stated above on the pdf and
on the process are verified (self similarity, $\alpha-stable$
process, stationary increments) but with a finite variance ($\alpha
> 2$) allowing to a unique value $H=1/2$.}. These results can be
compared with the study of the pdf $P(\Delta y,\Delta n=1)$ of the
positional jumps in the transverse direction, that showed in
contrary that these jumps were correlated in time for small imposed
deformation \cite{Tanguy2006}.

According to the above, a plausible  equation that would describe
the  evolution  within a L\'evy flight process for the stress
component $\sigma$, averaged over the whole system, is

\begin{equation}
\frac{{d(\sigma(i,\epsilon)-\overline\sigma(\epsilon))}}{d\epsilon}=\eta(i,\epsilon)
\label{dsigma}
\end{equation}
where $\eta(i,\epsilon)$ is a stochastic process whose spatial
average $\overline\eta(\epsilon)$ is the process that is entirely
characterized above (ie by the distribution of the elementary
increment and by the absence of correlation between successive
increments), and $\epsilon$ is the {\it external} imposed shear
strain. For small imposed shear strain, it can be mentioned that
such an equation with a noise corresponding to a L\'evy motion
cannot be reinterpreted in terms of the usual Fokker-Planck equation
as the moments of the shear stress $\sigma$ are non vanishing for
all orders. However, for large strain intervals $\Delta\epsilon$,
the stochastic process becomes gaussian, due to the existence of the
upper cut-off in the distribution of $\overline\eta$.

We will now propose a microscopic origin for the elementary process
$\eta$, by comparing it with a spatial redistribution of quadrupolar
stress.

But before, it can be interesting (as in
Ref.~\cite{Lemaitre2007,Lemaitre2007a}) to identify the contribution
to the distribution $P(\Delta\sigma,\Delta n)$ of events with a
release of the macroscopic stress (plastic events with
$\Delta\sigma_{macro}<0$). This is shown in figure 7(b). We see here
that for large imposed strains ($\Delta n > 64$ it means
$\Delta\epsilon > 0.32\%$ - smaller than the observed plastic
threshold) the contribution to the distribution is mainly due to
plastic events meaning that the weight of elastic events is
negligible in this range of applied strain. This corresponds also to
the strain interval $\Delta n$ at which the distribution becomes
gaussian with a good approximation (non gaussian parameter less than
1), and where the departure from L\'evy motion is of course
noticeable. For smaller strains, the contribution of steps with an
increase of the macroscopic stress ($\Delta\sigma_{macro}> 0$)
influences the distribution mainly for small values of the local
stresses. These positive steps are for example responsible for the
different power-law decay ($\propto 1/\Delta\sigma_{xy}^{2.5}$ in
place of $\propto 1/\Delta\sigma_{xy}^{1.7}$) in the beginning of
the first cross-over in the distribution (at the end of the initial
plateau). We are unable to explain the precise value of these two
exponents. However, we will now propose a model that is able to
explain the generic features of this distribution of shear stress
increments.

\section{Minimal model: sum of quadrupoles}

As shown in \cite{Tanguy2006}, the plastic flow regime consists in a
succession of elastic branches and plastic events. We will now focus
on the contribution of plastic events to the evolution equation of
the stress components in the system. We identify here a plastic
event by the fact that the associated stress is relaxed
macroscopically (corresponding to an evident dissipation of energy
shown by a negative slope in the stress-strain response of the
system, see Fig.1). Some of them correspond to an isolated
quadrupolar redistribution of local stresses around a well
identified center inside the sample; other plastic events consists
in an alignment of rotational structures in the particle
displacements giving rise to elementary shear bands going through
the system \cite{Tanguy2006}. The two kinds of events coexist in
the plastic flow regime, and the transition from one kind of plastic
event to another is still matter of debate
~\cite{Lemaitre2007,Marmottant2007,Marmottant2007a,Graner2007}.
However, the isolated quadrupolar event is more frequent than the
elementary shear band. In the linear initial regime for example,
only the isolated quadrupolar events take place, while there are
periods with only quadrupolar events in the plastic flow regime as
well.

In the following, we take the quadrupolar isolated event as the
elementary building block for explaining the plastic deformation of
the material, and we propose  a simple model that describes the
plastic deformation (as shown in the redistribution of stresses) as
a sum of uncorrelated quadrupoles of random amplitude $A$.

We first identify the distribution $P(A)$ in our system. As shown by
Picard et al.~\cite{G.2004}, a quadrupolar event involves in 2D a long-range
redistribution of stresses, due to a local pure shear. The
corresponding stress change is of the form.

\begin{equation}
\label{eqn:Dquadru}
\Delta\sigma_{xy}(r,\theta)\equiv \left\{
     \begin{array}{cc}
        A\frac{r_{0}^{2}}{r^{2}}  & \mbox{, if}      \  r_{0}\leq r\leq r_{max}  \\
        A                     & \mbox{, if}      \  0\leq r \leq r_{0}
     \end{array}
   \right\}
\end{equation} where $r_{0}$ and $r_{max}$ are
respectively the typical size of the quadrupole and the size of the
system, $A$ is the amplitude of the quadrupoles, and
$\Delta\sigma_{xy}$ denotes the incremental shear stress. We have
neglected in this expression the quadrupolar angular dependence of
the stress field and only considered its $\frac{1}{r^{2}}$ spatial
decay \footnote{The angular dependence will mainly contribute by a
scaling factor, and affect the weight of small $\Delta\sigma$ in a
logarithmic way in the final result. See M.Tsamados et al. {\it
preprint} (2007) for a detailed calculation.}. The distribution of
$P(A)$ is measured in our data (see figure 8). It corresponds to the
incremental stress at the center of the quadrupole, that is at the
place where the displacement is maximum. The best fit is
exponential, with $P(A) = \frac{1}{2\sigma_A}.\exp(-\frac{\vert
A\vert}{\sigma_A})$ and $\sigma_A = 2$ appears to be the
characteristic amplitude of the quadrupolar event. It is independent
on the system size, unlike the distribution of macroscopic stress
release that is $\propto 1/V$ (figure 2), and is symmetric.

Using the exponential distribution $P(A)$ shown above, we can now
reproduce the distribution $P(\Delta\sigma=s,\Delta n=1)$ of the
incremental shear stress $\Delta\sigma_{xy}$ averaged over the whole
system. According to the previous analysis (see part III), the
distribution for $\Delta n > 1$ should then be reproduced by
successive convolutions, assuming the absence of temporal
correlations. Neglecting the angular dependence of
$\Delta\sigma_{xy}$ as in Eq. \ref{eqn:Dquadru}, we show that, for a
given amplitude $A$, there is a simple bijective relation between
the shear stress $\Delta\sigma_{xy}$ and the radial coordinate $r$.
We can thus write $P_A(\Delta\sigma_{xy})d\Delta\sigma_{xy} = P(r)
dr$, with $P(r) dr = 2\pi r / V dr$. Using Eq.\ref{eqn:Dquadru}, we
get
\begin{equation}
P_A(\Delta\sigma_{xy}) = \frac{\pi A r_{0}^2}{V\Delta\sigma^2}
\end{equation}
with
\begin{equation}
\label{eqn:amp}
\vert A\vert\frac{r_{0}^2}{r_{max}^2} \le \vert\Delta\sigma_{xy}\vert \le \vert A\vert
\end{equation}
In the following, we will consider only the case $A > 0$, the
opposite case giving the symmetric distribution corresponding to
$\Delta\sigma_{xy} < 0$. Assuming that the successive events
contribute independently to the total distribution
$P(\Delta\sigma_{xy})$ (an assumption justified by the fact that
this distribution is averaged over the time origins), the average
probability of having an incremental stress $\Delta\sigma_{xy}$
inside the system is thus obtained by summing over all quadrupolar
events
\begin{equation}
P(\Delta\sigma_{xy}) = c\int_{\Delta\sigma_{xy}}^{\Delta\sigma_{xy}\frac{r_{max}^2}{r_{0}}} P_A(\Delta\sigma_{xy})P(A)dA
\end{equation}
with an additional normalisation factor $c$ due to the limited range of allowed amplitudes.
This gives
\begin{eqnarray}
P&(&\Delta\sigma_{xy}) = c\frac{\pi r_0^2}{2V}
\frac{1}{\Delta\sigma_{xy}^2}((\sigma_A+\Delta\sigma_{xy})exp(-\Delta\sigma_{xy}/\sigma_A)
\nonumber\\
 - &(&\sigma_A+\sigma_{xy}\frac{r_{max}^2}{r_0^2})exp(-\Delta\sigma_{xy}.r_{max}^2/(\sigma_A.r_0^2)))
\end{eqnarray}
We show here that the upper exponential cut-off is proportional to
$\exp(-\Delta\sigma_{xy}/\sigma_A)$ and that the low $\Delta\sigma$
behavior\footnote{with a logarithmic divergence if the angular
dependence is taken into account see M. Tsamados et al. {\it
preprint} (2007).} is dominated by finite size scaling proportional to
$\exp(-\Delta\sigma_{xy}.r_{max}^2/(\sigma_A.r_0^2))$. We have
plotted this function in figure 9, and compared it with the result
obtained numerically. We find a good agreement with the numerical
result, for $\sigma_A = 2$, $r_{max} = L = 100$ and $r_{0} = 4$. The
parameter $r_{0}$ is the unique free parameter of the fit since
$r_{max}$ scales like the system size and the fitted value for
$\sigma_A$ corresponds to the value obtained in the measured
distribution of $P(A)$ (see figure 8), thus confirming our simple model. The small
value obtained for $r_{0}$ shows that the plastic quadrupolar events
are localized.

This simple model gives us a physical intuition on the origin of the
distribution for the increments $\overline\eta$ observed in the
averaged evolution equation \ref{dsigma}. Therefore we can now make
more precise the evolution equation at a  local level as

\begin{equation}
\frac{\partial (\sigma(\overrightarrow{r},
\epsilon)-\overline\sigma(\epsilon))}{\partial \epsilon}=\int d^3\overrightarrow{r_Q}\frac{A(\epsilon)
 r_{0}^2}{\vert\vert \overrightarrow{r}-\overrightarrow{r_{Q}(\epsilon)}
 \vert\vert^{2}}\rho(\overrightarrow{r_Q},\epsilon)
\end{equation}
where $\overrightarrow{r_{Q}}$ is the position where the quadrupolar
event has occurred,$\vert\vert \overrightarrow{r}-\overrightarrow{r_{Q}}
\vert\vert \geq r_{0}$ , and $\rho(\overrightarrow{r_Q},\epsilon)$ is the
number of plastic event per unit strain and unit volume, taking place at $\overrightarrow{r_{Q}}$ when the external strain is $\epsilon$.

As it appears clearly here, this equation is not complete since
$\overrightarrow{r_Q}(\epsilon)$ is unknown. In many models
~\cite{Picard2005,Baret2002}) the occurrence of a quadrupolar event
at position $\overrightarrow{r_{Q}}$ is triggered by a local stress
threshold. In the following section, we discuss the existence of
such a criterion.

\section{Looking for a local threshold}

The center of the quadrupolar plastic rearrangements is identified
as the place where the particle displacement is the largest~\cite{Tanguy2006}.
In order to relate the position of the center of the quadrupoles to
local stress criteria, we have compared the distribution of the
stress components obtained for the whole system, and the value of
the same stress component on the particle in the center of the
quadrupole, one step before the quadrupole takes place.

The result for the deviatoric stress
$\sigma_{dev}\equiv\sigma_1-\sigma_2$ is shown in figure 10(a). The
distribution of the deviatoric stress in the whole sample is
stationary: it is the same in the beginning (for ex. one step before
the first plastic event) and in the end (e.g. one step before the
last measured plastic event) of the plastic flow. It is remarkable
to see in this figure, that the distribution of the deviatoric
stress restricted to the center of the quadrupoles (one step before
each plastic event) computed for all the plastic events is very
close to the previously discussed stationary distribution obtained
over the whole sample, and not only in the center of the
quadrupoles. The position of the center of the quadrupoles appears
to be unrelated with any threshold value in the local stress
components. The saturation of the macroscopic stress, and the
associated well defined yield stress $\sigma_Y$ in these systems can
thus be related only to the alternance of increase and decrease of
stress and to its intermittent behavior, rather than to any
identified local stress threshold. The existence of an apparent
macroscopic yield stress is not related to a local yield stress.
Note that criterion based on Tresca or Mohr-Coulomb criterion,
involving the most probable relation between deviatoric stress and
pressure have also been tested in our numerical systems. It shows a
general tendency for the pressure to be affinely related to the
deviatoric stress in the plastic flow regime. However, this tendency
is shared by all the particles in the system, and not only by the
particles at the center of the quadrupoles. It thus appears, like in
the case of the deviatoric stresses, that a global Mohr-Coulomb
criterion, or equivalently -for 2D systems- a Tresca criterion, is
valid in the plastic flow regime, but is does not provide  a locally
selective criterion for plasticity.

Figure 10(b) shows the same distributions as for the deviatoric stress,
but for the incremental deviatoric stresses:
it means $\sigma_{dev}(n+1)-\sigma_{dev}(n)$.
In this case, quenched stresses are not taken into account, and a
large incremental stress is generally the signature of a large
local deformation. The distribution of incremental stresses is also
stationary. The distribution restricted to the center of the
quadrupoles is displaced to larger increases in the deviatoric
stresses. This point has also been mentioned by Robbins et al.
~\cite{Rottler2003} and underlines the role played by incremental
stresses in the dissipative dynamics of the systems, in comparison
with total stresses that are more or less unchanged.
It suggests that the incremental stress $\Delta\sigma$ can be a more
important parameter for plastic purposes than the stress $\sigma$ itself.
However, its distribution is very broad, and a criterion based only on
the incremental stresses is not very selective: it does not
even exclude the possibility for the center of the quadrupole to
take place where the increase of the deviatoric stress is not
maximum but minimum...

The question of the existence of a local criterion for plasticity
appears thus clearly beyond our detailed numerical study of local
stress components, and  deserves a further studies.

\section{Conclusion}

We have shown in this paper that the local stress components evolve
in a heterogeneous way during quasi-static plastic deformation. It
is related to a finite cooperativity number identifying fluctuating
zones whose maximum size is approximately $6\times6$ particle
diameters in the plastic flow regime. The strain at which these
zones form ($\approx 4\%$) is related to the typical strain at which
the memory of the local stress seems to be lost in the 2-point
correlation function.

It is thus natural to propose a picture of the heterogeneous stress
evolution, as a sum of localized plastic events. We have shown in
this paper that the averaged evolution of the full distribution of
local stress components can be described by a simple model. Within
this model, the evolution of local stress components results from
the uncorrelated sum of localized quadrupolar rearrangements. The
model allows to recover the 3 regimes shown in the distribution: its
$1/V$ dependence at small stresses, in agreement with a local
description of the plastic instability, the cross-over to an
apparent power-low decay $\propto 1/\Delta\sigma^{\alpha+1}$ with
$\alpha$ close to $1$ in the intermediate range, and for larger
stresses a size-independent upper cut-off directly related in our
model to the size-independent distribution of stress rearrangements
in the center of the quadrupoles.

However, a complete description of local stress rearrangements is
still lacking: for example, the observed exponential nature of the
distribution of the amplitudes of stress rearrangements in the
center of the elementary quadrupoles - a crucial ingredient of our
model - is still not explained, and the position of the center of
the quadrupoles should be related to a criterion of instability not
identified in this paper. This last question especially deserves
further studies, since is appears clearly here that the local
stresses are not determinent.

Any attempt to describe local plasticity in terms of local stress
rearrangements should thus take into account these numerical results
before proposing a complete description of the heterogeneous
mechanical response in terms of a single evolution equation.

\bf{Acknowledgments} During the course of this work, we have
benefited of discussions from L.~Bocquet, M.~Robbins, S.~Roux,
P.~Sollich, and D.~Vandembroucq. Computational support by
IDRIS/France, CINES/France and CEA/France is acknowledged.

\bibliographystyle{epj}
\bibliography{biblio_short}

\onecolumn
\newpage
\begin{figure}[!hbtp]
\begin{center}
\includegraphics*[width = \textwidth]{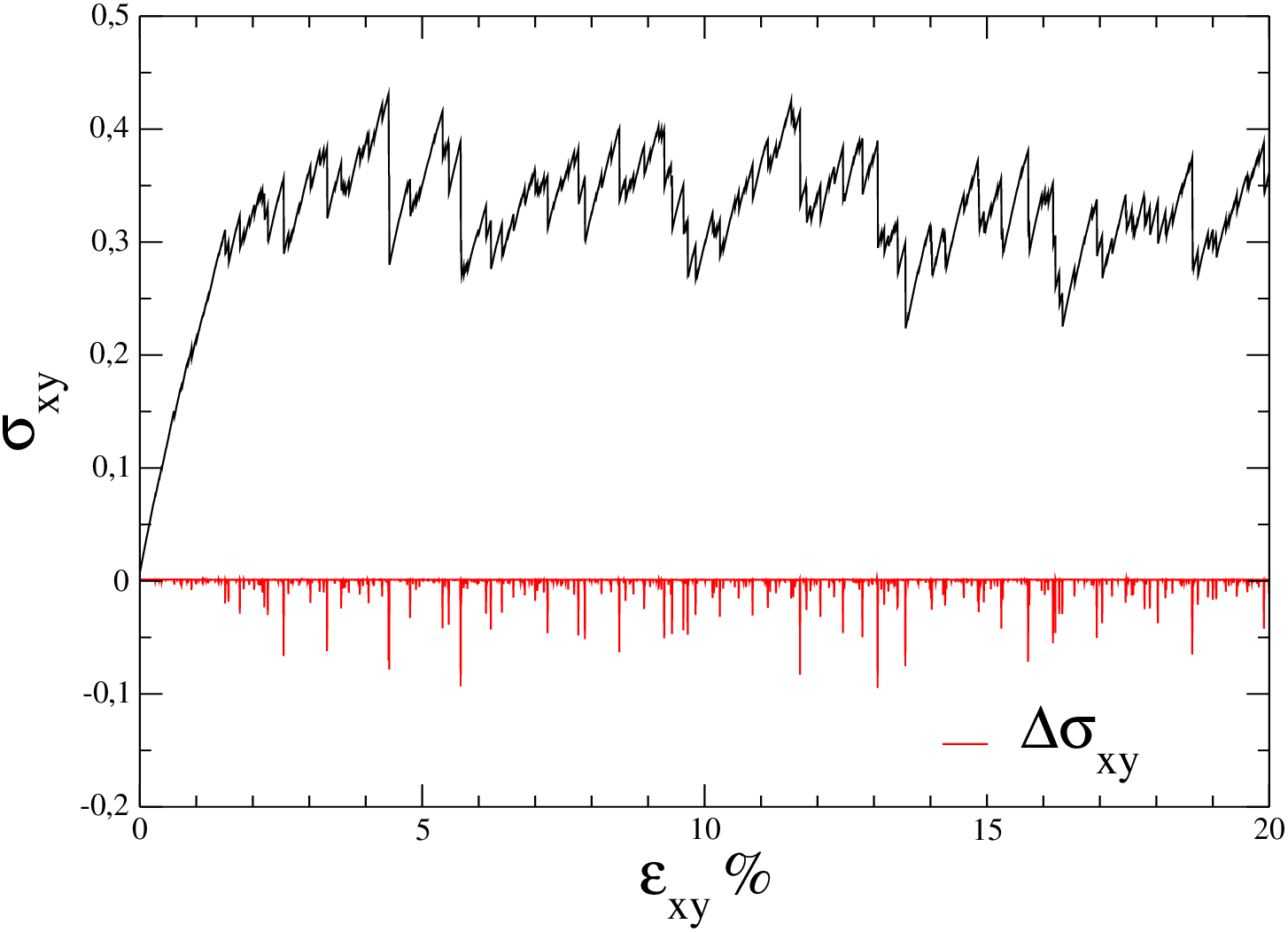}
\caption[stress-strain]{stress-strain response and incremental shear
stress,for one configuraiton of 10 000 particles.}
\label{fig:stress-strain}
\end{center}
\end{figure}

\newpage
\begin{figure}[!hbtp]
\begin{center}
\includegraphics*[width = \textwidth]{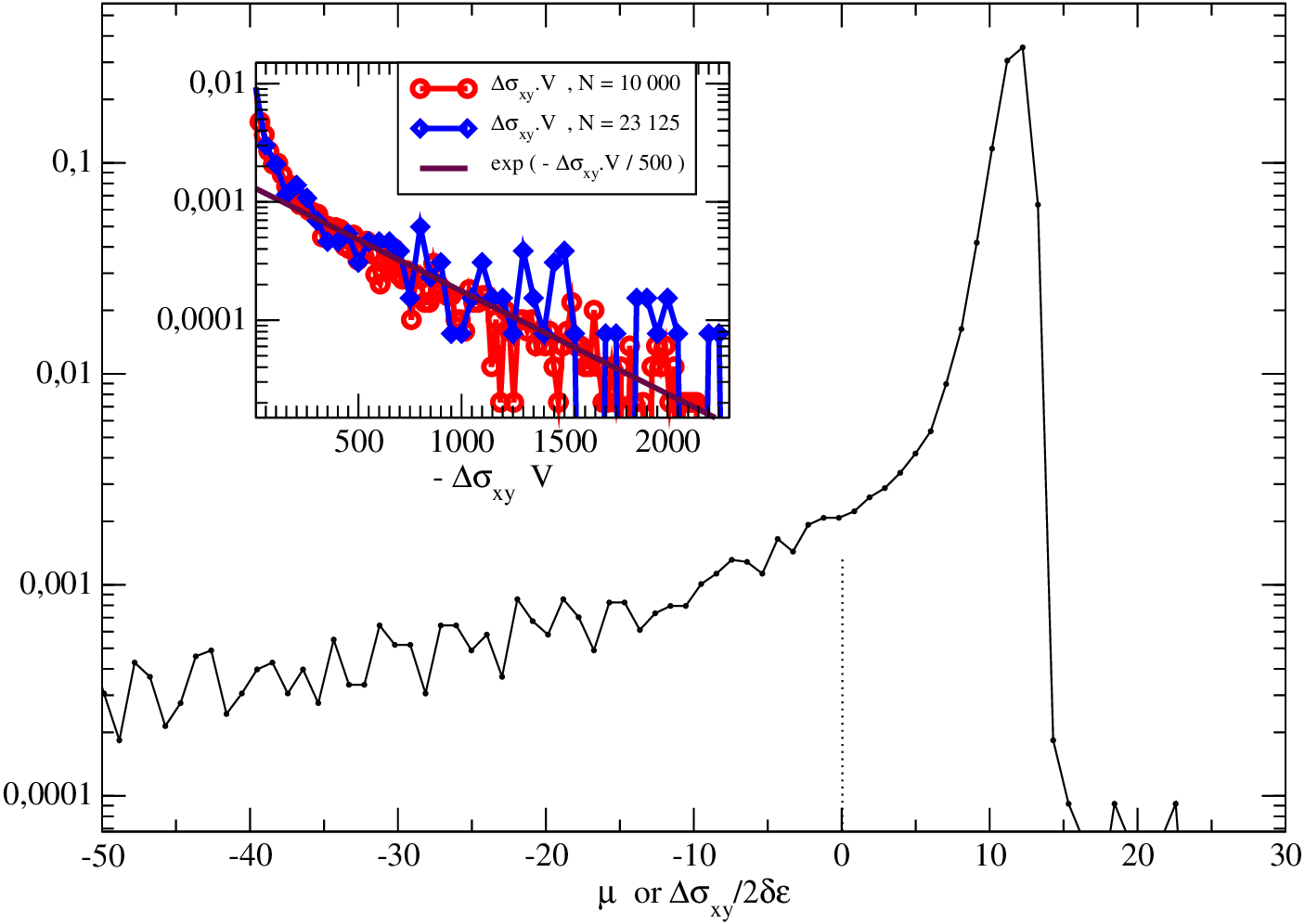}
\caption[mu-distrib]{Histogram of the slope
$\Delta\sigma/\Delta\epsilon$ of the macroscopic shear stress.
Inset: Size dependence of the amplitude of negative stress changes
$\Delta\sigma_{xy}$. The amplitude of stress releases is
proportional to the volume of the sample. } \label{fig:mu-distrib}
\end{center}
\end{figure}

\newpage
\begin{figure}[!hbtp]
\begin{center}
\includegraphics*[width = \textwidth]{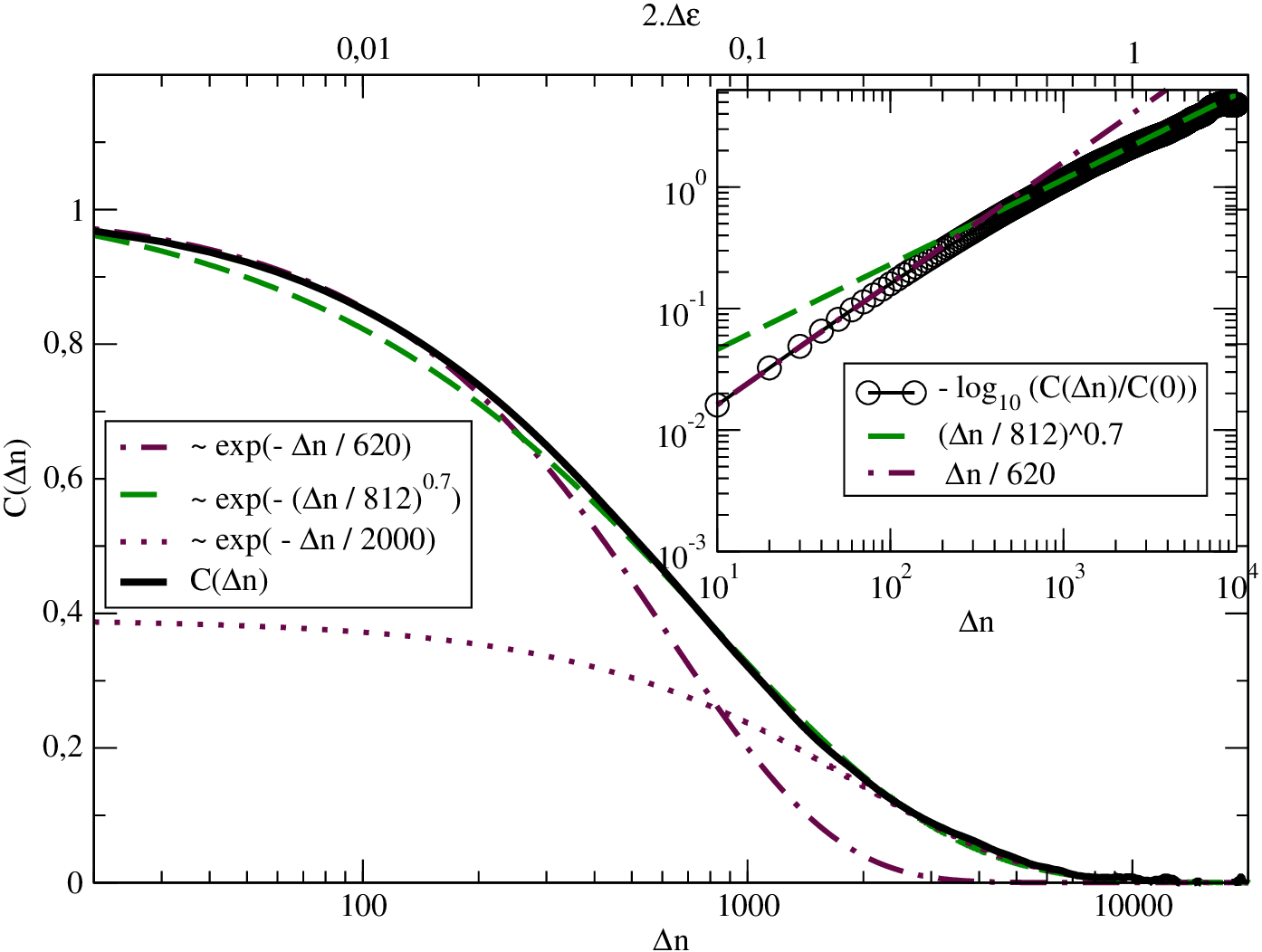}
\caption[corr-temp]{Autocorrelation function of the shear stress
computed on each particle and averaged over the whole sample, as a
function of time (applied shear strain). The nature of the fits is
indicated in the legend box. Inset: Logarithm of the Autocorrelation
function of the shear stress, in log-log scale, in order to
determine the exponent of the corresponding stretched exponential}
\label{fig:corr-temp}
\end{center}
\end{figure}

\newpage
\begin{figure}[!hbtp]
\begin{center}
\includegraphics*[width = 0.8\textwidth]{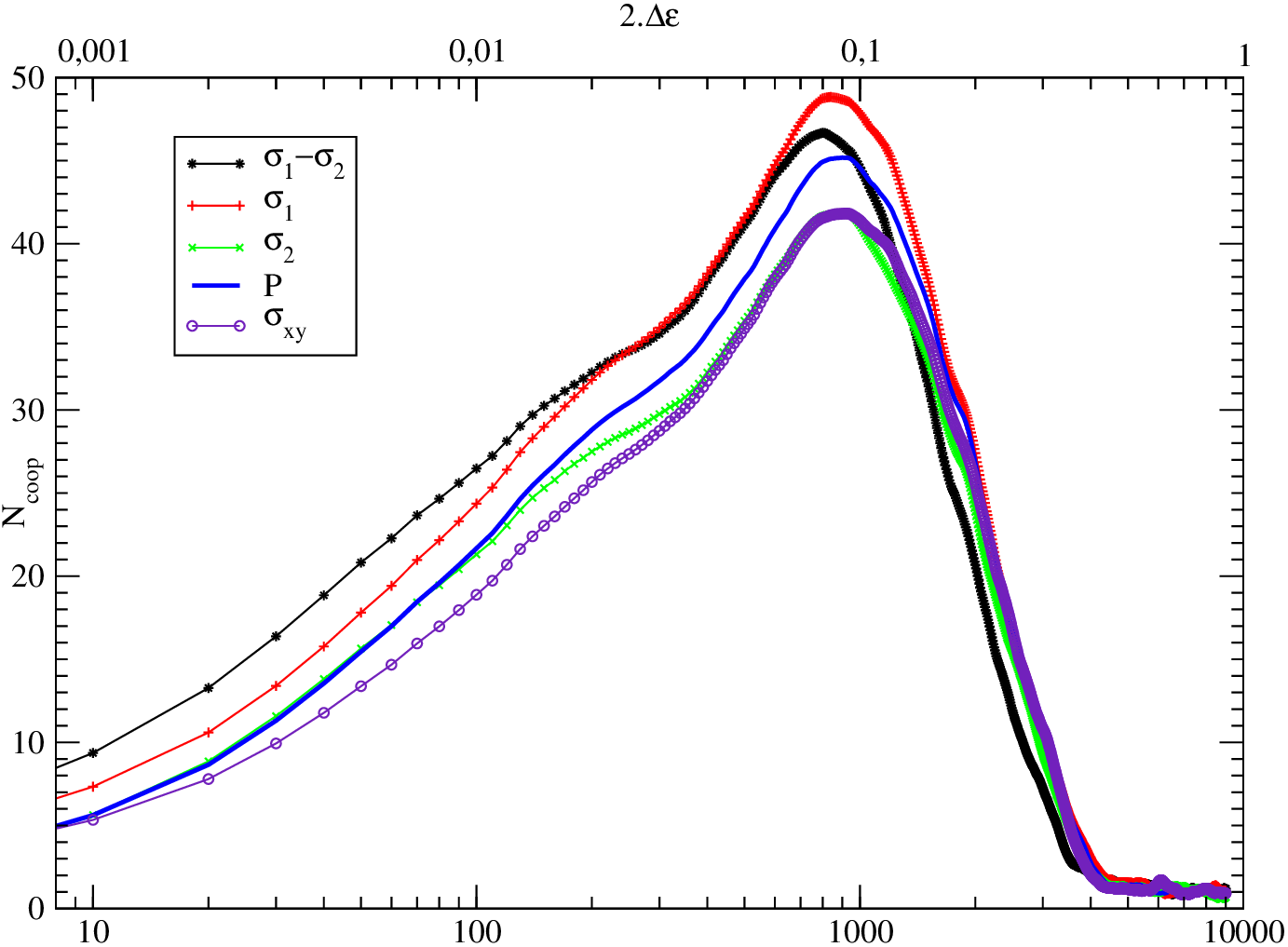}
\includegraphics*[width = 0.8\textwidth]{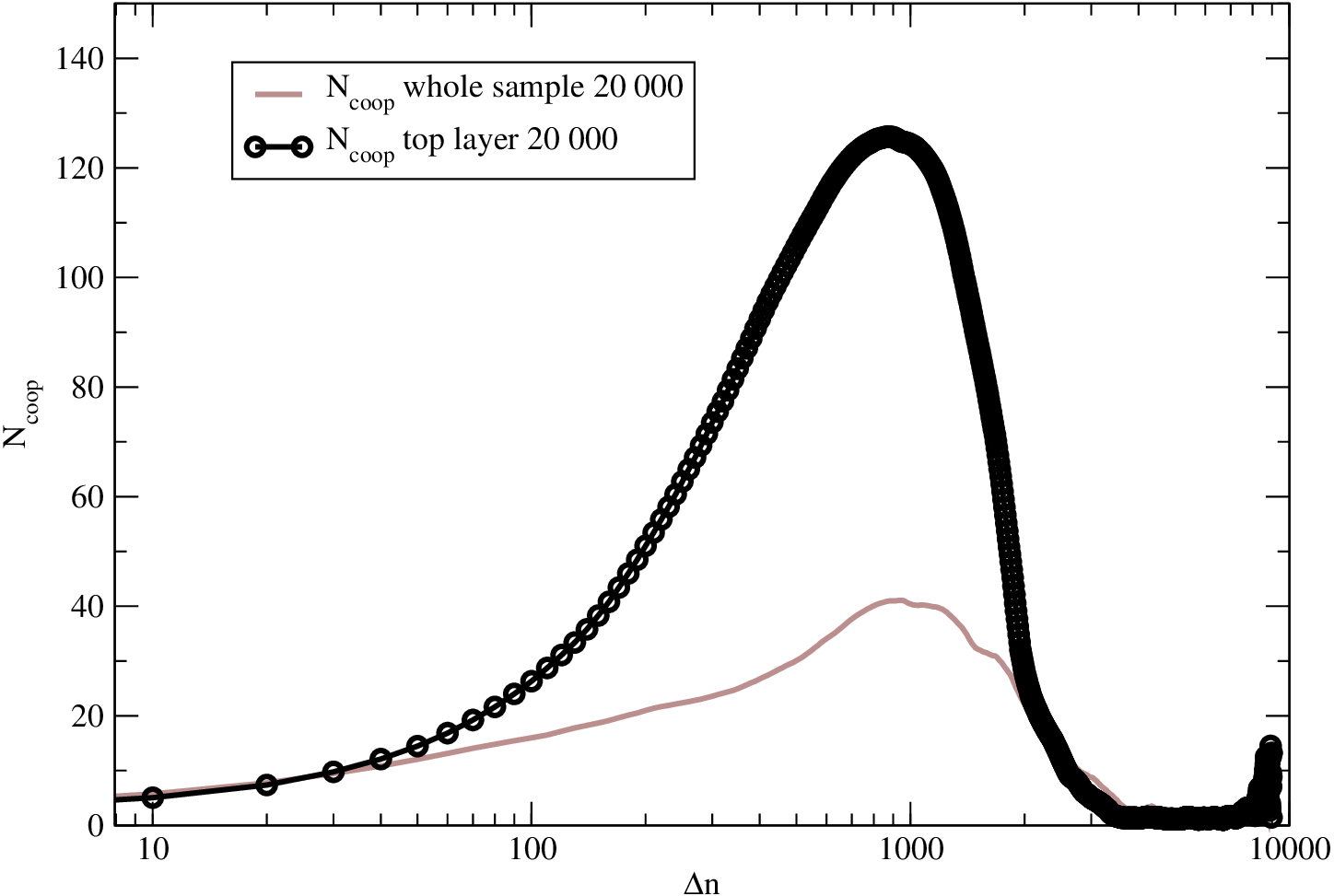}
\caption[Khi4]{(a) Cooperativity number as a function of time
(applied shear strain) for various stress components: local shear
stress $\sigma_{xy}$, local pressure $P$, eigenvalues of the stress
tensor $\sigma_1$ and $\sigma_2$, and the deviatoric stress
$\sigma_{dev}=\sigma_1-\sigma_2$ (from bottom to top).
The curves have been obtained by averaging on 20 000
time origins $t_0$.
(b) Cooperativity number of the local shear stress at the
borders, in comparison with the cooperativity number in the center
of the sample} \label{fig:Khi4}
\end{center}
\end{figure}

\newpage
\begin{figure}[!hbtp]
\begin{center}
\includegraphics*[width = \textwidth]{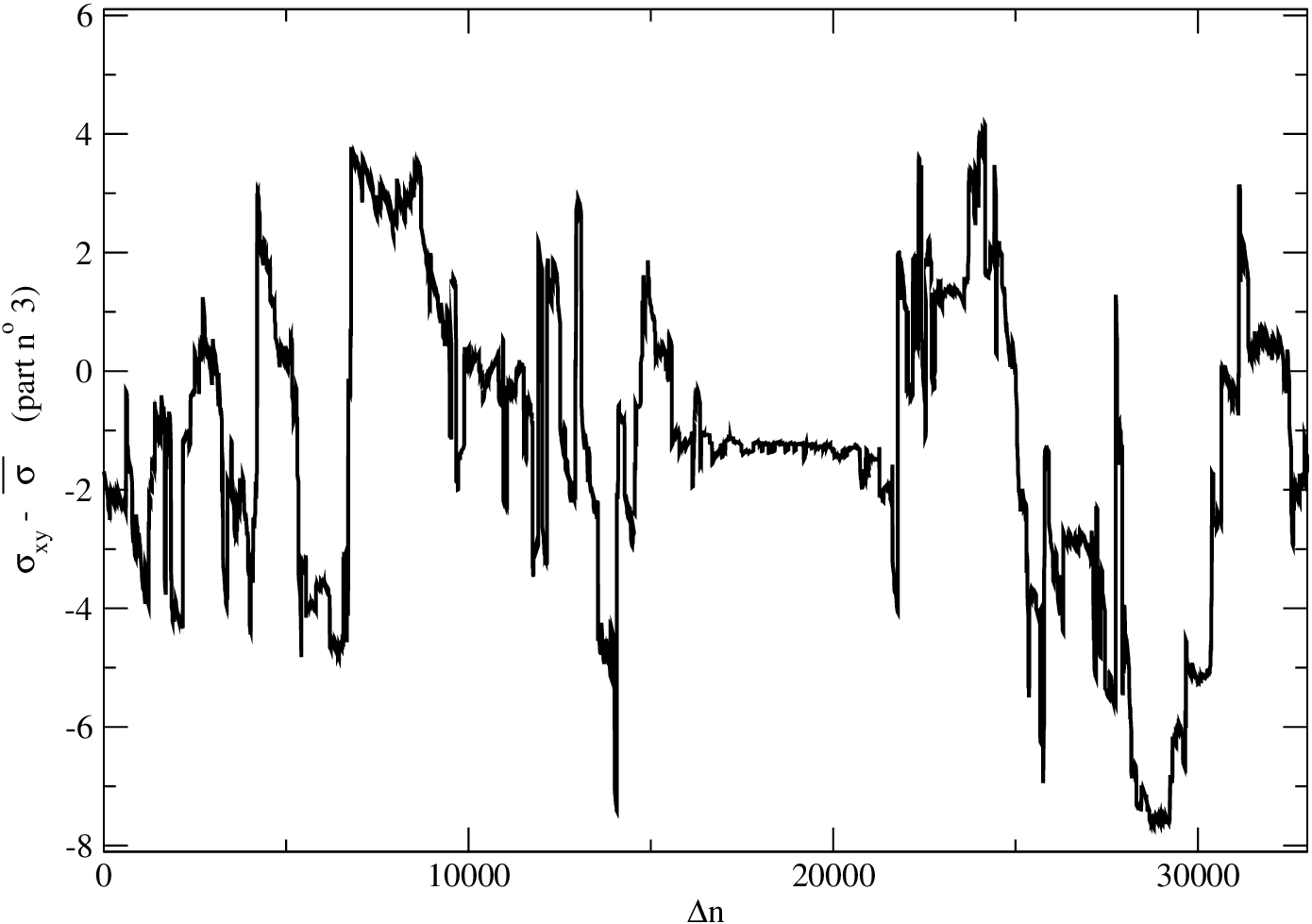}
\caption[stress-part-i]{Evolution of the shear stress on a given
particle in the center of the sample, as a function of the applied
shear strain. For each value, the average over the sample has been
substracted.} \label{fig:stress-part-i}
\end{center}
\end{figure}

\newpage
\begin{figure}[!hbtp]
\begin{center}
\includegraphics*[width = 0.5\textwidth]{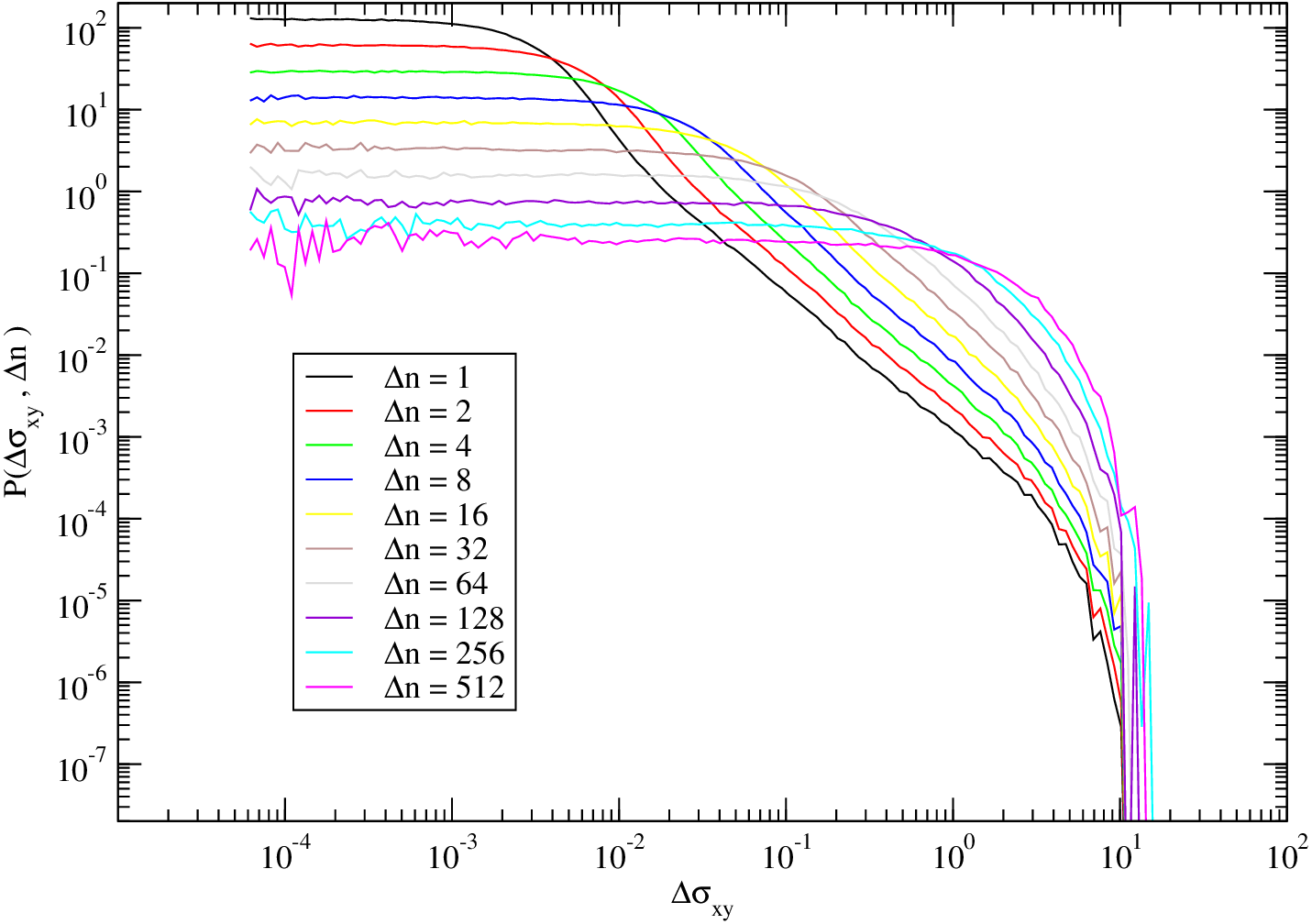}
\includegraphics*[width = 0.5\textwidth]{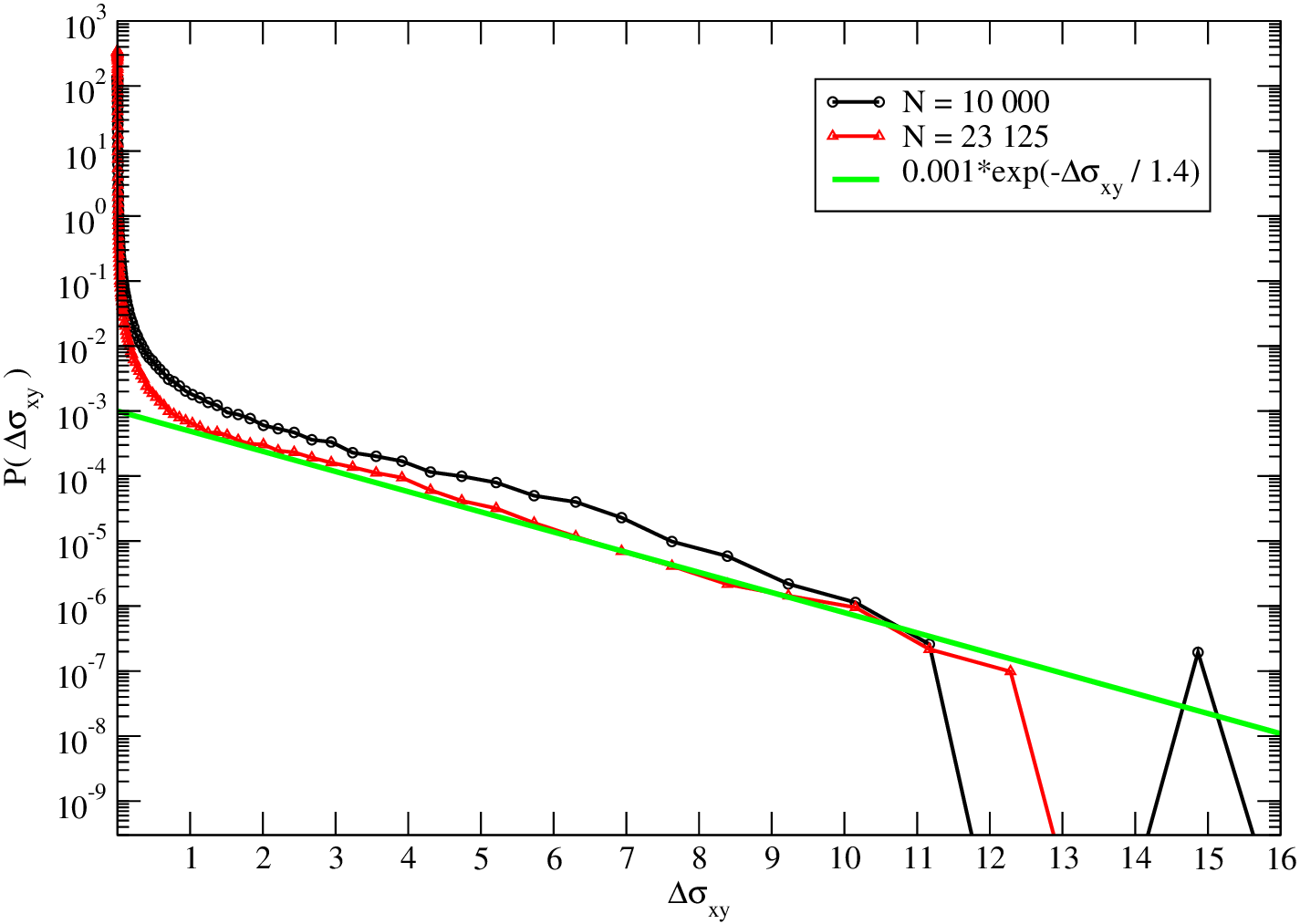}
\includegraphics*[width = 0.5\textwidth]{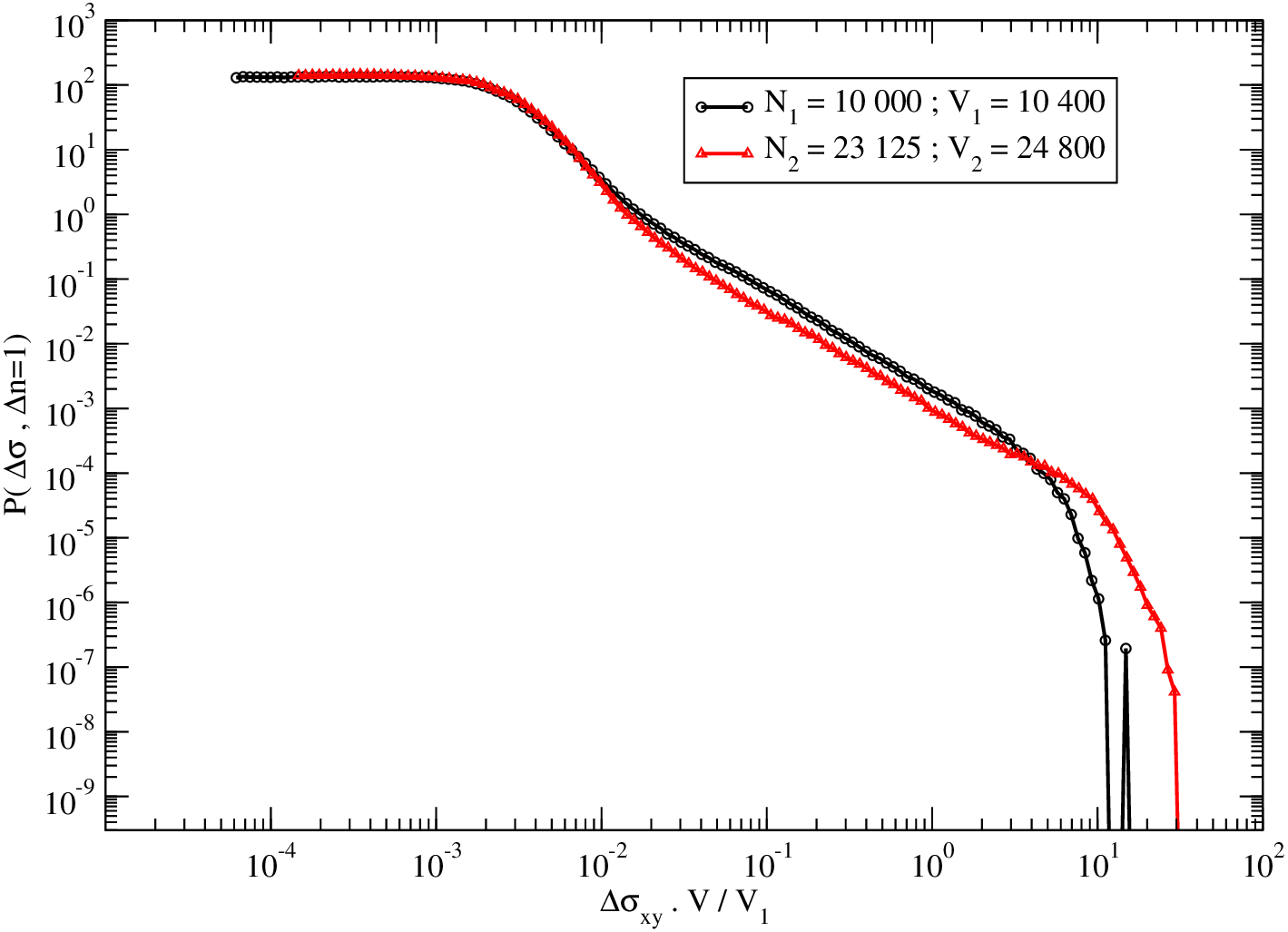}
\caption[stress-distrib]{(a) distribution
$P(\Delta\sigma_{xy},\Delta n)$ of the variation of local shear
stress $\Delta\sigma_{xy}$ for different strain intervals $\Delta
n$. The distribution has been obtained by averaging over the whole
sample, and for different time origins. (b) lin-log scale for
$P(\Delta\sigma_{xy},\Delta n = 1)$ and different system sizes $N =
10 000  (104\times 104)$ and $N = 23 125 (50\times 500)$. It shows a
size independent exponential upper cut-off with a characteristic
$\Delta\sigma \approx 1.4 $. (c) log-log scale for
$P(\Delta\sigma_{xy},\Delta n = 1)$ and different system sizes $N =
10 000 (104\times 104)$ and $N = 23 125 (50\times 500)$. It shows
the scaling of the first cross-over $\propto 1/V$ where $V$ is the
volume of the system. The stress variation is mainly due to local
variation that averages over the whole system (see text). }
\label{fig:stress-distrib}
\end{center}
\end{figure}

\newpage
\begin{figure}[!hbtp]
\begin{center}
\includegraphics*[width = 0.8\textwidth]{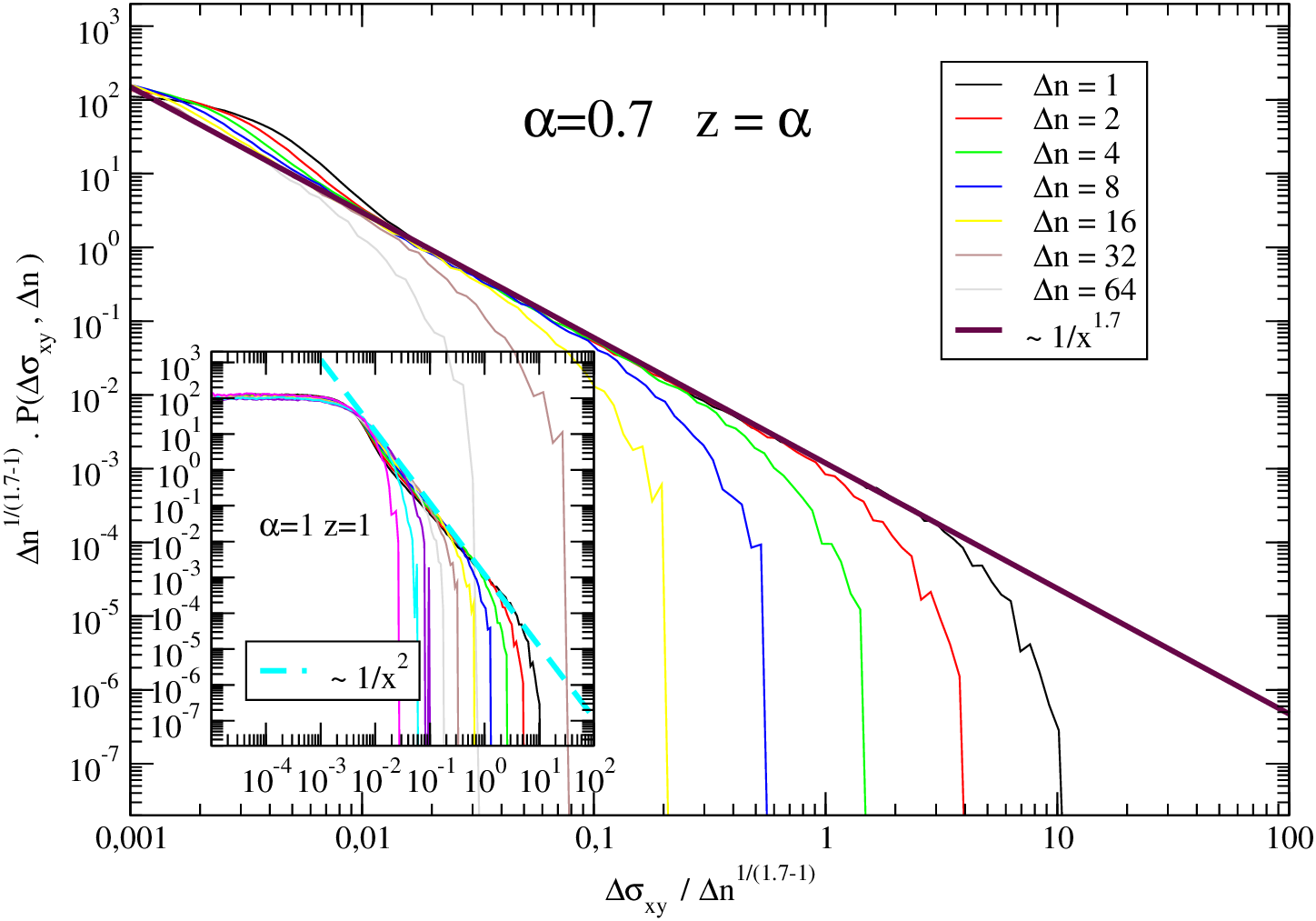}
\includegraphics*[width = 0.8\textwidth]{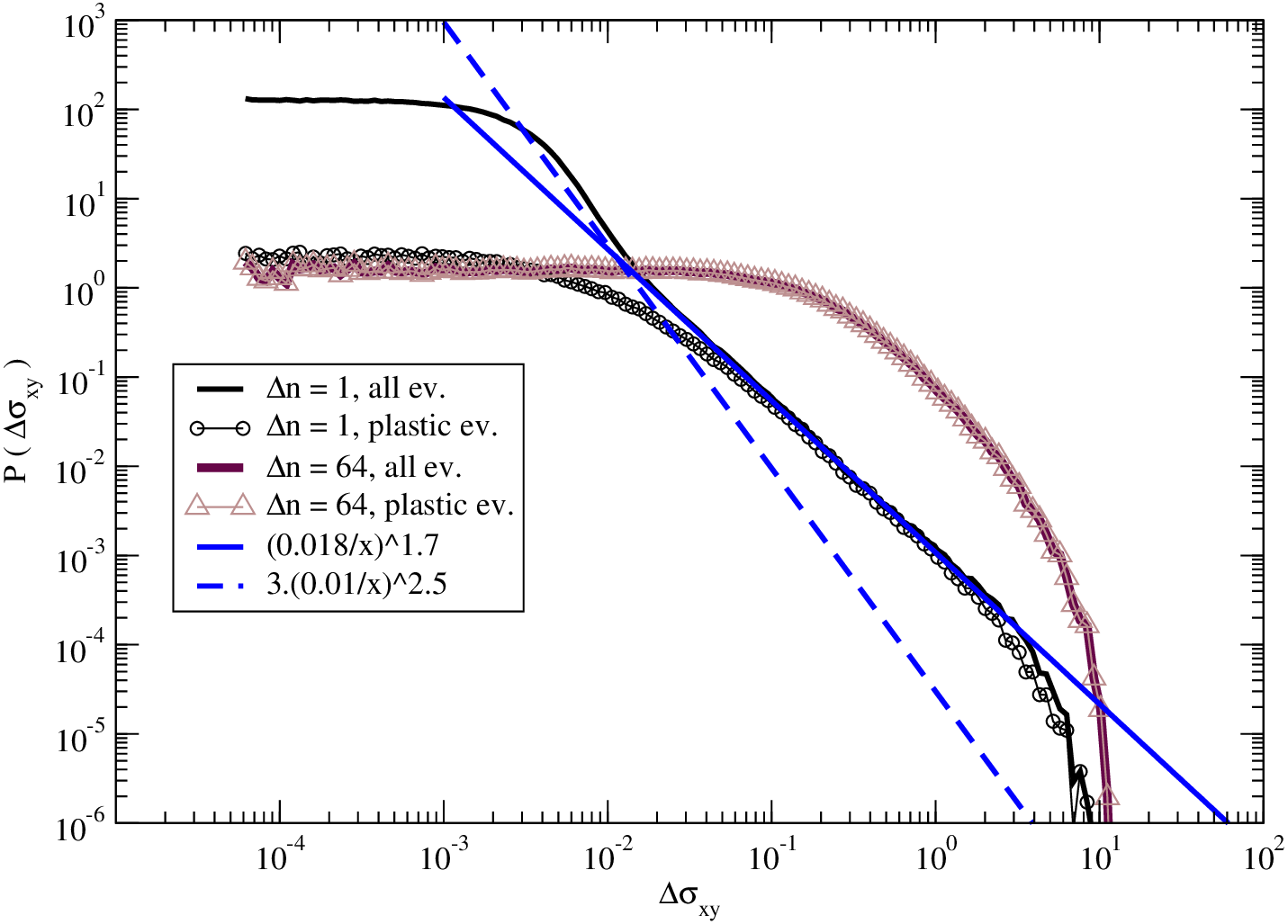}
\caption[stress-scaling]{(a) scaling of $P(\Delta\sigma_{xy},\Delta
n)\propto\phi(\Delta\sigma_{xy}/\Delta n^{1/z})$ with $z=\alpha$ and
$\alpha=0.7$. The scaling is very good in the intermediate range of
the distribution, and for small $\Delta n < 64$ where the
exponential cut-off does not play a role. Inset: the same with
$\alpha=1=z$. This exponent corresponds also to the one discussed in
the text. (b) separate contribution of steps associated with a
release of stress at a macroscopic level, to the total distribution
of local stresses, for time intervals $\Delta n=1$ and $\Delta n =
64$}. As shown here, the distribution of $\Delta\sigma$ for $\Delta
n \geq 64$ is dominated by the contribution of stress changes due to
plastic events. \label{fig:stress-scaling}
\end{center}
\end{figure}

\newpage
\begin{figure}[!hbtp]
\begin{center}
\includegraphics*[width = \textwidth]{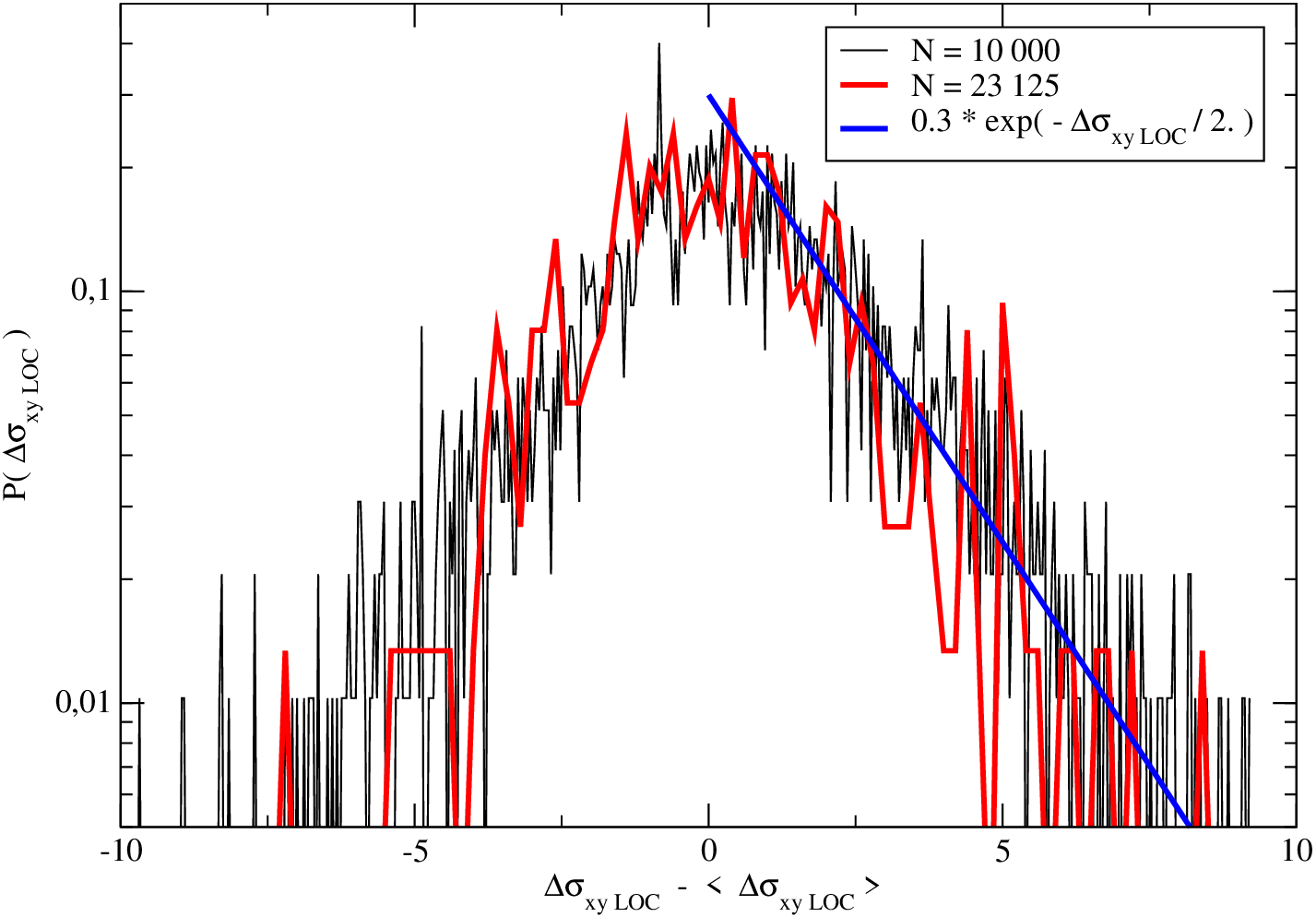}
\caption[stress-micro-thresh]{Distribution of shear stress releases
in the center of the quadrupoles, during a plastic event. The fit is
exponential with a characteristic $\Delta\sigma = 2$. }
\label{fig:stress-micro-thresh}
\end{center}
\end{figure}

\newpage
\begin{figure}[!hbtp]
\begin{center}
\includegraphics*[width = \textwidth]{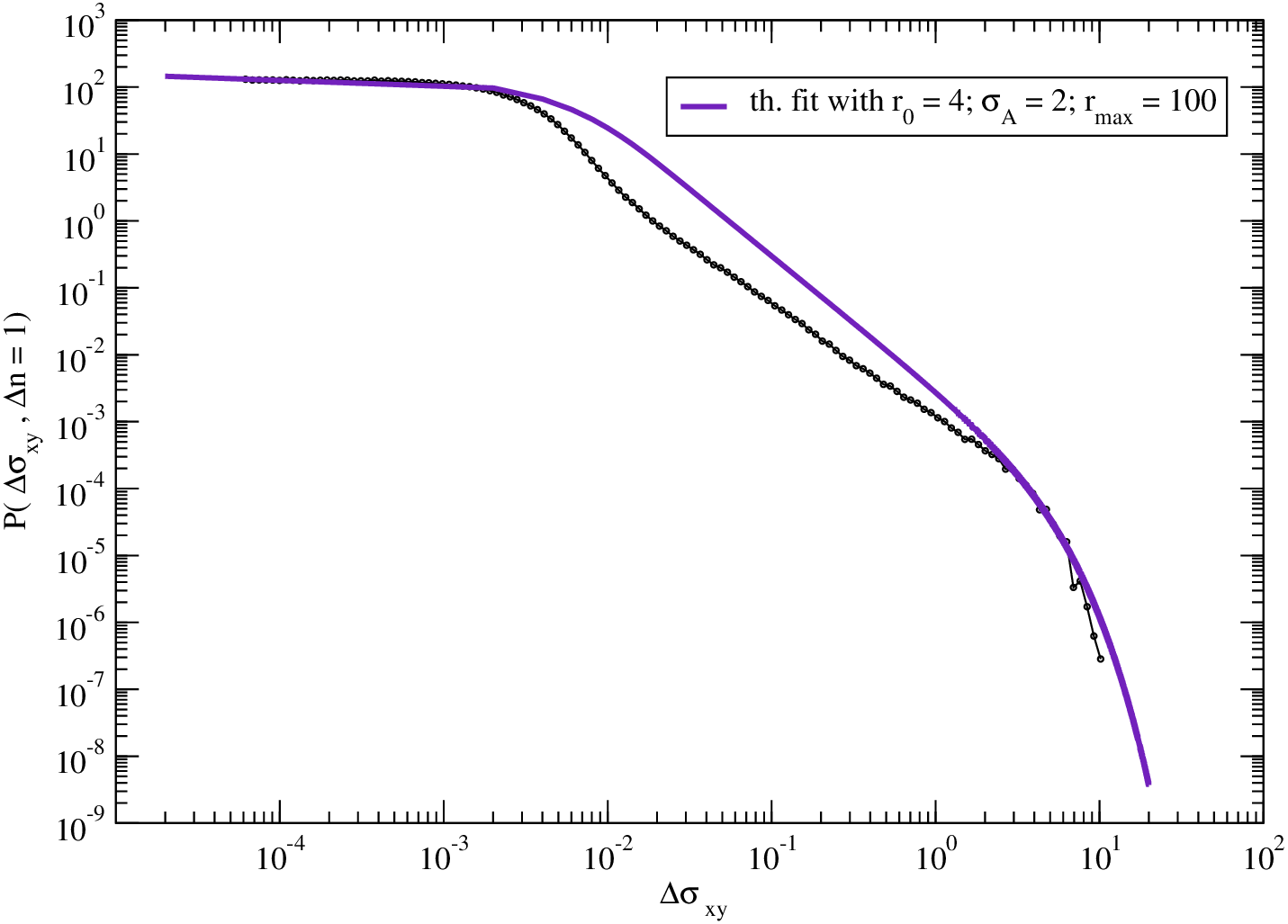}
\caption[stress-fit]{Comparison of the distribution
$P(\Delta\sigma_{xy},\Delta n=1)$ with the theoretical fit discussed
in the text. The parameters of the fit are indicated in the legend
box.} \label{fig:stress-fit}
\end{center}
\end{figure}

\newpage
\begin{figure}[!hbtp]
\begin{center}
\includegraphics*[width = 0.8\textwidth]{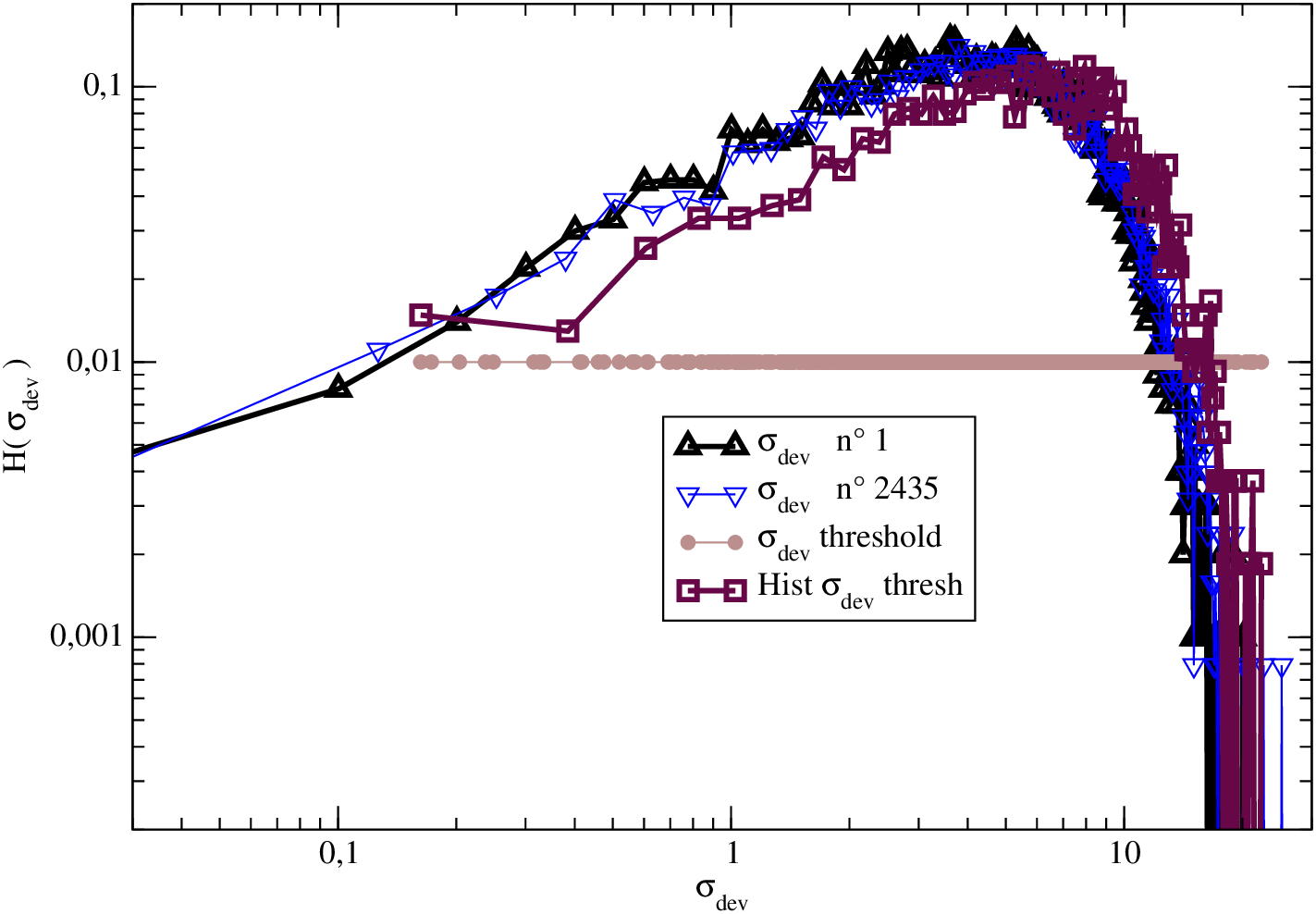}
\includegraphics*[width = 0.8\textwidth]{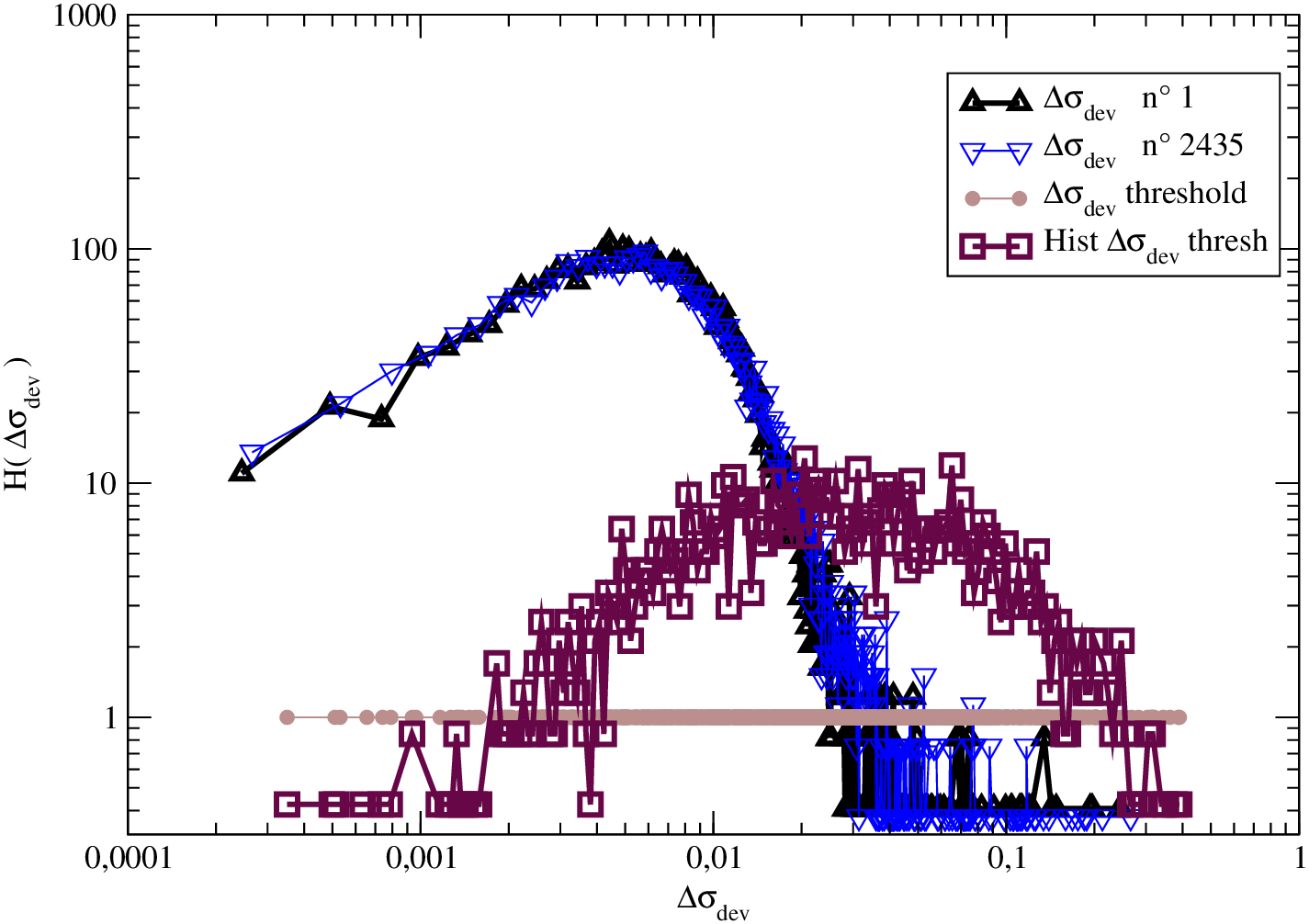}
\caption[Histo-stress]{(a) Distribution of the
deviatoric stress in a
configuration of 10 000 particles for 2 different strains in the
beginning and in the end of the plastic flow plateau (triangles).
Comparison with the distribution of the deviatoric stress in the
center of the future quadrupole just before the quadrupolar event
takes place (squares), computed on the 2500 plastic events of the
plastic flow regime. The 3 curves are very similar. (b) Same as in
(a), but for the {\it incremental} deviatoric stress. In this case,
the distribution obtained on the center of the quadrupoles indicates
a more pronounced correspondance with the highest incremental
stresses inside the sample.} \label{fig:Histo-stress}
\end{center}
\end{figure}

\end{document}